\documentclass[preprint2]{aastex}
\usepackage{amsmath}
\usepackage{natbib}
\usepackage{footnote}
\usepackage{lscape}

\begin{document}


\title{Combining Photometry From \emph{Kepler} and TESS To Improve Short-Period Exoplanet Characterization}


\author{Ben Placek\altaffilmark{2} and Kevin H. Knuth \altaffilmark{1}}
\affil{Physics Department, University at Albany (SUNY),
    Albany, NY 12205 \\
placekbh@sunysccc.edu, kknuth@albany.edu}
\author{Daniel Angerhausen}
\affil{NASA Goddard Space Flight Center, Exoplanets \& Stellar Astrophysics Laboratory, Code 667, Greenbelt, MD 20771\\
daniel.angerhausen@nasa.gov}

\altaffiltext{1}{Department of Informatics, University at Albany (SUNY), Albany, NY 12205}
\altaffiltext{2}{Center for Science and Technology, Schenectady County Community College, Schenectady NY, 12305}

\begin{abstract}
Planets emit thermal radiation and reflect incident light that they recieve from their host stars.  As a planet orbits it's host star the photometric variations associated with these two effects produce very similar phase curves.  If observed through only a single bandpass this leads to a degeneracy between certain planetary parameters that hinder the precise characterization of such planets.  However, observing the same planet through two different bandpasses gives one much more information about the planet.  Here, we develop a Bayesian methodology for combining photometry from both \emph{Kepler} and the Transiting Exoplanet Survey Satellite (TESS). In addition, we demonstrate via simulations that one can disentangle the reflected and thermally emitted light from the atmosphere of a hot-Jupiter as well as more precisely constrain both the geometric albedo and dayside temperature of the planet. This methodology can further be employed using various combinations of photometry from the James Webb Space Telescope (JWST), the Characterizing ExOplanet Satellite (CHEOPS), or the PLATO mission.
\end{abstract}



\section{Introduction}
With the advent of high-precision photometry a hidden component to the light curve of exoplanets was revealed: the phase curve.  The photometric phase curve corresponding to an exoplanet consists of four known effects: the reflection of incident stellar light \citep{Jenkins&Doyle:2003,Seager+etal:2000,Perryman:2011,Placek+etal:2014}, and the emission of thermal radiation from the atmosphere (or surface) of the planet \citep{Charbonneau:2005, Borucki+etal:2009,Placek+etal:2014}, the relativistic Doppler beaming of star-light as the host star orbits the system's center of mass \citep{Rybicki&Lightman:1979, Loeb&Gaudi:2003,Placek+etal:2014}, and the tidal warping of the stellar surface due to the proximity of the massive planet known as ellipsoidal variations \citep{Loeb&Gaudi:2003, Placek+etal:2014}.  Each effect has been observed for short-period hot-Jupiters in the \emph{Kepler} field \citep{Esteves+etal:2013, Angerhausen&Delarme:2014, Borucki+etal:2009, Faigler&Mazeh:2011, Faigler+etal:2013, Placek+etal:2014, Lillo-Box+etal:2013, Mazeh+etal:2011, Mislis&Hodgkin:2012, Shporer+etal:2011, Welsh+etal:2010}. While the Doppler beaming and tidal warping effects provide information on the planetary mass, reflection and thermal emission yield information on the planetary radius, geometric albedo, and day-side temperature. 

In the case of circular orbits, the reflection and thermal emission components to the phase curve are largely indistinguishable --- they both manifest as sinusoidal modulations.  \citet{Placek+etal:2014} showed that it should be possible to disentangle thermal and reflected photons for sufficiently close-in eccentric orbits if the day-side temperature of the planet does not change significantly during apastron passage. 

The amount of thermal flux recieved from an exoplanet depends on the bandpass through which the planet is observed.  A significant degeneracy exists between the geometric albedo and dayside temperature when a light curve is analyzed in only a single bandpass.  However, when two (potentially overlapping) bandpasses are employed and the geometric albedo of the planet does not change significantly between bandpasses, it is possible to break this degeneracy and significantly improve the characterization of such planets.   \citet{Shporer+etal:2014} used a similar approach with the transiting exoplanet \emph{Kepler}-13Ab in which the planetary system was observed in three separate bandpasses --- \emph{Kepler}, IRAC/3.6$\mu$m, and IRAC/4.5 $\mu$m.  This allowed them to greatly constrain both the albedo and day-side temperature.

The Transiting Exoplanet Survey Satellite (TESS) will be conducting an all-sky survey of transiting exoplanets and thus providing the opportunity to further characterize short-period hot-Jupiters when it observes the \emph{Kepler} field \citep{Ricker+etal:2015}.  TESS will observe the nearest and brightest stars with magnitudes of $+4 < V < +12$ at a short 2 minute cadence, and will provide full-frame images at a cadence of 30 minutes allowing for the study of even dimmer stars in each of the observation sectors.  According to the Exoplanet Orbit Database \citep{Han+etal:2014}, there are 37 confirmed planetary systems in the \emph{Kepler} field around stars brighter than $V = 12$ and an additional 47 systems around stars brighter than $V = 15$, which is the expected limit of the TESS full frame images.  Of these 84 systems, 20 are thought to be short period planets with orbital periods less than 15 days, and of the planets for which there is a published $M_p\sin i$, there are $13$ hot-Jupiters. These planets and their published characteristics are listed in Table \ref{Known Planets}. Currently the K2 mission is observing stars brighter than those observed by \emph{Kepler} ($V < 12$) at a expected photometric precision of $\approx 80$ ppm for a 6-hour integration and 1-minute cadence \citep{Howell+etal:2014}, increasing the number of known planets that can be further characterized with TESS.

Here, we present a study of model-generated photometric data for a transiting short-period hot-Jupiter observed through both the \emph{Kepler} and TESS bandpasses.  This synthetic planet is based off of published parameter estimates from the exoplanet Kepler13Ab, which orbits a hot A-type star in the \emph{Kepler} field.   The host star has a magnitude of V = 9.9, making it a great candidate to characterize with both \emph{Kepler} and TESS.  We employ Bayesian model selection to show that one can disentangle thermal and reflected light when both bandpasses are simultaneously analyzed and provide parameter estimates displaying a breaking of the albedo vs. day-side temperature degeneracy. In addition, we quantify the extent to which using multiple bandpasses can aid in exoplanet characterization and provide a framework for similar studies for future missions.  

\begin{table*}
\centering
\begin{tabular}{lccccc}
Name 			&	Orbital Period (d)					&	Planetary Mass ($M_J$)	& 	Stellar Mass ($M_\odot$)				&	V-Magnitude	\\
\hline
TrES-2b			&	$2.47063 \pm 1.0\times10^{-5}$		&	$1.201 \pm 0.052$		&	$0.983^{+0.063}_{-0.059}$	&	$11.4$		\\
HAT-P-7 			& 	$2.204737 \pm 1.7\times10^{-5}$        &	$1.792 \pm 0.063$		&	$1.50 \pm 0.03$			&	$10.5$		\\
\emph{Kepler}-12b	&	$4.43796370 \pm 2.0 \times 10^{-7}$	&	$0.432 \pm 0.042$		&	$1.166^{+0.051}_{-0.054}$	&	$13.8$		\\
\emph{Kepler}-13Ab&	$1.76358799 \pm 3.7 \times10^{-7}$	&	$6.5 \pm 1.57$		&	$1.72 \pm 0.1$			&	$9.9$		\\
\emph{Kepler}-14b 	&   	$6.7901230 \pm 4.3\times10^{-6}$	&	$8.41 \pm 0.29$		&	$1.512 \pm 0.043$			&	$12.0$		\\
\emph{Kepler}-17b	&	$1.48571080 \pm 2.0\times10^{-7}$	&	$2.48 \pm 0.102$		&	$1.16 \pm 0.06$			&	$13.8$		\\
\emph{Kepler}-40b	&	$6.87349 \pm 6.4 \times19^{-4}$		&	$2.18 \pm 0.34$		&	$1.48 \pm 0.06$			&	$14.8$		\\
\emph{Kepler}-41b	&	$1.8555580 \pm 7.1\times10^{-6}$	&	$0.494 \pm 0.071$		&	$0.94 \pm 0.09$			&	$14.5$		\\
\emph{Kepler}-74b	&	$7.3407180 \pm 1.0\times10^{-6}$	&	$0.667 \pm 0.09$		& 	$1.40^{+0.14}_{-0.11}$		&	$14.2$		\\
\emph{Kepler}-412b &	$1.720891232 \pm 4.7\times10^{-8}$	&	$0.940 \pm 0.087$		&	$1.167 \pm 0.091$			&	$13.7$		\\
\emph{Kepler}-423b &	$2.684328480	\pm 8.2\times10^{-8}$	&	$0.723 \pm 0.100$		&	$1.07 \pm 0.05$			&	$14.5$		\\
\emph{Kepler}-424b &	$3.31186440\pm3.9\times10^{-7}$	&	$0.77 \pm 0.23$		&	$1.010 \pm 0.054$			&	$14.5$		\\
\emph{Kepler}-425b &	$3.79701816 \pm 1.9\times10^{-7}$	&	$0.249 \pm 0.074$				&	$0.93 \pm 0.05$			&	$15$		\\
\hline

\end{tabular}
\caption{Thirteen of the brightest known \emph{Kepler} planets with periods less than fifteen days.  All data were retrieved from the Exoplanet Orbit Database.}\label{Known Planets} 
\end{table*}

\section{Bayesian Inference} \label{Bayes'}
Bayes' Theorem is used to make inferences from data and is given by
\begin{equation}
P(\theta_M | D, M) = \frac{P(\theta_M|M)P(D|\theta_M,M)}{P(D|M)}
\end{equation}
where $D$ represents a given dataset, $M$ a particular model described by a set of model parameters $\theta_M$.  The prior probability, $P(\theta_M | M)$, quantifies one's knowledge of the parameter values $\theta_M$ before analyzing the dataset $D$.  Bayes' takes the data, $D$, into account through the likelihood function $P(D|\theta_M,M)$, which represents the probability that one would observe the data given the set of model parameters $\theta_M$ of the model $M$.  The denominator is a normalization constant known as the Bayesian evidence, or marginal likelihood,  which quantifies the probability that the model $M$ could have produced the data $D$.  Together, these three quantities yield the posterior probabilty $P(\theta_M|D,M)$, which quantifies one's knowledge of the model parameters after the data has been analyzed.  

Therefore, Bayes' Theorem acts as an updating rule for probabilities: one begins with their prior knowledge about a particular system, observes data about the system, and ends up with the posterior probability after having analyzed the data via the likelihood function.  The posterior probability is essential in obtaining parameter estimates through summary statistics.  The Bayesian evidence is more important for determining to what extent a model $M$ describes the observed data. 

\subsection{Bayesian Model Selection}

\begin{table*}[t!]
\begin{center}
\begin{tabular}{l c c l}
  Parameter & Variable & & Distribution \\
\hline
Dayside Temp. (K) & $T_d$  &  & U(0,6000) \\
Nightside Temp. (K) & $T_n$ & & U(0,6000) \\
Orbital Inclination &  $\cos i$ & & U(0,1) \\
Planetary Radius $(R_J)$ & $R_p$& & U(0,3) \\
Planetary Mass $(M_J)$ & $M_p $ & & U(0,10) \\
Geometric Albedo & $A_{g}$ & & U(0,1) \\
Standard Dev. of noise (\emph{Kepler}, ppm) & $\log \sigma_K $ & & U(-15,0) \\
Standard Dev. of noise (TESS, ppm) & $\log \sigma_T $  & & U(-15,0)\\
\hline
 \end{tabular}
\end{center}
\caption{Prior distributions for planetary and orbital parameters where U signifies a uniform distribution over the range inside the parentheses.}\label{Priors}
\end{table*} 

Bayesian Model Selection relies on the Bayesian evidence, $P(D|M)$, which is commonly denoted $Z$, and is calculated by normalizing the posterior probability in Bayes' Theorem:
\begin{equation}
Z = \int P(\theta_M|M) P(D,|\theta_M,M) d\,\theta_M
\end{equation}
This integration is performed over each model parameter, which typically makes this integral analytically intractable and requires one to employ numerical integration techniques. It can be shown using Bayes' theorem that the ratio of posterior probabilities between two competing models $M_1$ and $M_2$ with equal prior probabilities, is equal to the ratio of the evidence for each model.  Therefore the model with the larger evidence value will be considered to be more favorable to describe the data \citep{Knuth+etal:2014}.  The amount by which one model with evidence $Z_1$ is favored over another model with evidence $Z_2$ is typically quantified using the Bayes' Factor, which is given by
\begin{equation}
O = \frac{Z_1}{Z_2}.
\end{equation}
Often one focuses on the logarithm of the Bayes' factor, which is given by
\begin{equation}
\ln O = \ln{Z_{1}} - \ln{Z_{2}}.
\end{equation}
Guidelines for interpreting Bayes' Factors were given both by \citet{Jeffreys:1939} and \citet{Kass&Raftery:1995}.  There is overwhelming evidence for a particular model being favored over a competing model if the log-Bayes' factor, $\ln O$, is greater than five, strong evidence if it is between $2.5 < \ln O < 5.0$, positive evidence if between $1.0 < \ln O < 2.0$, and hardly significant below one \citep{vonderLinden+etal:2014}.  

 Since this integral is performed over the entire parameter space, there is an inherent Occam's razor effect.  Complex models typically have large parameter spaces where the probability of the model being correct is spread out over a larger region of parameter values.  Alternatively, simpler models with smaller parameter spaces have the probability spread out over a smaller region of parameter space.  Therefore, if both a simple and complex model describe the data equally well, the simple model will have a higher evidence and thus be the preferred model.

In order to compute model evidences the MultiNest algorithm \citep{Feroz&Hobson:2008, Feroz+etal:2009, Feroz+etal:2013}, which is a variant of the Nested Sampling algorithm \citep{Sivia&Skilling:2006}, was utilized.  Given a model, prior probability assignments, and a likelihood function, MultiNest provides estimates of the Bayesian log-evidence for a particular model as well as posterior samples from which parameter estimation can be performed.

\subsection{Priors and Likelihood Function}
As described in Section \ref{Bayes'}, in order to make inferences using Bayes' Theorem, one needs to assign prior probabilities for each model parameter.  For this study, each model parameter is assigned a uniform prior probability over a reasonable range as shown in Table \ref{Priors}.  The uniform prior probability assignment for the log of the signal variance is equivalent to Jeffreys Prior, which is an uninformative prior for scale parameters \citep{Jeffreys:1946,Sivia&Skilling:2006} .  Each of these assignments can always be modified to incorporate additional information.

 Since there are two datasets in this study, there must be two likelihood functions, one for the \emph{Kepler} time series ($L_{\text{Kep}}$) and another for TESS ($L_{\text{TESS}}$).  The joint probability for observing a datum $D$, given a set of model parameters $\theta_m$ corresponding to a model $M$ can be written as the product of the two likelihood functions
\begin{equation}
P(D|\theta_m,M) = L_{\text{Kep}} L_{\text{TESS}}
\end{equation}
In practice, due to the magnitude of the likelihood functions, it is common to use the log-likelihood function so that
\begin{equation}
\log P(D|\theta_m,M) = \log L_{\text{Kep}} + \log L_{\text{TESS}}
\end{equation}
where $\log L_{\text{Kep}}$ and $\log L_{\text{TESS}}$ independently take the observed data from \emph{Kepler} and TESS into account, respectively.  These two log-likelihood functions are chosen to be Gaussian based on the expected nature of the noise in both observations.  However, this can be changed to account for a variety of situations such as data with apparent correlated (red) noise \citep{Sivia&Skilling:2006}.  When the mean value of the signal and the signal variance are the only relevant parameters, it can be shown by the principle of maximum entropy that a Gaussian likelihood is the least biased choice \citep{Sivia&Skilling:2006} of likelihood function.  This yields a log-likelihood function of the form
\begin{align} \label{logL}
\log L = & -\frac{\chi_{\text{Kep}}^2}{2} - \frac{N_{\text{Kep}}}{2} \log 2\pi \sigma_{\text{Kep}}^2 \nonumber\\
	      & -\frac{\chi_{\text{TESS}}^2}{2} - \frac{N_{\text{TESS}}}{2} \log 2\pi \sigma_{\text{TESS}}^2 
\end{align} 
where $N_{\text{Kep}}$, $\sigma_{\text{Kep}}$, $N_{\text{TESS}}$,  and $\sigma_{\text{TESS}}$ are the total number of data points and expected squared deviation of the noise in the \emph{Kepler} and TESS datasets, respectively. The model-dependent $\chi^2$ terms are given by 
\begin{equation}\label{chi2}
\chi^2 = \sum_{i = 1}^N \left( F(t_i) - D_i \right)^2
\end{equation} 
where $F(t_i)$ represents the forward model evaluated at the times $t_i$, which the data were observed.

\subsection{Forward Model}

The likelihood function depends explicitly on the forward model $F(t_i)$ as shown in (\ref{logL}) and (\ref{chi2}).   The forward model computes the observed flux originating from a planetary system modeled by the parameters $\theta_M$ of model $M$.  In addition to transits and secondary eclipses, there are four photometric effects that must also be taken into account.  These include the reflection of incident stellar light off of the planetary surface and/or atmosphere, thermal emission from the planetary surface and/or atmosphere, relativistic Doppler beaming of light as the host star orbits the center of mass, and the ellipsoidal variations caused by tidal forces induced by the planet on the host star.  The stellar-normalized reflected light component of the flux is given by

\begin{equation}
\frac{F_R(t)}{F_\star} = \frac{A_g}{\pi} \frac{{R_p}^2}{r(t)^2}\left(\sin\theta(t) + (\pi - \theta(t)) \cos
\theta(t) \right)
\end{equation}
where $A_g$ is the geometric albedo, $R_p$ is the planetary radius, $r(t)$ is the planet-star separation distance, and $\theta(t)$ is the angle between the vector connecting the planet and star and the line of sight. 
The thermal emission from both the day- and night-side of the planetary atmosphere is given by 
\begin{equation}
\frac{F_{D}(t)}{F_{\star}} = \frac{1}{2} \frac{B(T_d)}{B(T_{eff})}\left( \frac{R_p}{R_s} \right)^2 \left( 1 + cos\theta (t)\right)
\end{equation}
and
\begin{equation}
\frac{F_{N}(t)}{F_{\star}} = \frac{1}{2} \frac{B(T_n)}{B(T_{eff})}\left( \frac{R_p}{R_s} \right)^2 \left( 1 + cos\theta (t)\right)
\end{equation}
where $R_\star$ is the stellar radius, and $B(T)$ is the Planck's law, which is evaluated at the day-side temperature of the planet $T_d$, and the effective temperature $T_{\text{eff}}$ of the host star.  The beaming effect is given by 
\begin{equation}\label{beaming}
\frac{F_{B}(t)}{F_\star} =  (3-\alpha_b)\beta_r(t)
\end{equation}
where $\beta_r(t)$ is the radial velocity of the host star and directly depends on the radial velocity semi-amplitude
\begin{equation}\label{semi-amplitude}
K = 28.435\left(\frac{T}{1yr} \right)^{-\frac{1}{3}} \frac{M_p\sin i}{M_J}\left(\frac{M_\star}{M_\odot}\right)^{-\frac{2}{3}},
\end{equation}
and $\alpha_b$ is the photon-weighted bandpass-integrated beaming factor given by
\begin{equation}\label{beaming_2}
\alpha_b = \frac{\int K(\lambda)B\lambda F_{\lambda,\star} d\,\lambda}{\int K(\lambda) \lambda F_{\lambda,\star} d\,\lambda}.
\end{equation}
Here, $K(\lambda)$ is the \emph{Kepler} response function, $\lambda$ is the wavelength, $F_{\lambda,\star}$ is the stellar spectrum, and $B$ is given by
\begin{equation}\label{beaming_3}
B = 5 + \frac{d \ln F_{\lambda,\star}}{d \ln \lambda}
\end{equation}
where the derivative is averaged over the observed wavelengths. The tidal effect is approximated as
\begin{equation}
\frac{F_{E}(t)}{F_{\star}} = \alpha \frac{M_p}{M_\star} \left( \frac{R_\star}{r(t)} \right)^3 \sin^2i \cos2\theta(t)
\end{equation}
where $M_p$ and $M_\star$ are the planetary and stellar masses, respectively, and $i$ is the orbital inclination.  The coefficient $\alpha$ is given by
\begin{equation}
\alpha = \frac{0.15(15+u)(1+g)}{3-u}
\end{equation}
 where $u$ and $g$ are the linear limb-darkening coefficient and gravity darkening coefficient for the host star and can be determined from the tables in \citet{Claret&Bloeman:2011} .

\begin{table*}
\centering

\begin{tabular}{cccc}
		 & 	\emph{Kepler} Only RBE		 &TESS Only RBE		&	\emph{Kepler} + TESS RBE 	\\
 \hline
 Parameter  	& Mean  					 & Mean 				&	Mean\\
\hline
$\cos i$	   	& $0.117 \pm 0.010$	 	&$0.117 \pm 0.010$	&$0.117 \pm0.010$		 \\
$R_p$	  	& $1.414 \pm 0.017$		&$1.408 \pm 0.015$	&$1.409 \pm 0.015$		\\
$M_p$	  	&$7.48 \pm 0.17	$		&$7.36 \pm 1.00$		&$7.36 \pm 0.96$				\\
$A_g$	  	&$0.58 \pm 0.01	$		&$0.56 \pm 0.072$		&$0.82\pm 0.07$			\\
$T_d$	  	&  \nodata				&\nodata				&\nodata				 \\
$T_n$	  	&\nodata					&\nodata				&\nodata					\\
$\sigma_K$    & $104.0 \pm 6.0$			&$108.0 \pm 6.0$		&$106.0 \pm 6.0$			\\
$\sigma_T$    &\nodata					&\nodata			        &$206.0 \pm 13.0$			\\
$\log Z$ & $500.9 \pm 0.4$   		 	&$1004.3 \pm 0.3$		&$949.0 \pm 0.2$			   \\
$\log L_{\text{max}}$ &  $516.1$				&$1021.6$			&	$969.9$					\\
\hline
			& \emph{Kepler} Only RBET	&TESS Only RBET		& \emph{Kepler} + TESS RBET\\
\hline
$\cos i$		&  $0.117 \pm 0.010$		&$0.116 \pm 0.015$	&$0.117 \pm 0.010$\\
$R_p$		&  $1.414 \pm 0.017$		&$1.409 \pm 0.016$	&$1.409 \pm 0.015$\\
$M_p$		&  $7.44 \pm 0.18$			&$7.44 \pm 1.08$		&$7.26 \pm 0.92$\\
$A_g$		&$0.46 \pm 0.16$			&$0.54 \pm 0.13$		&$0.27 \pm 0.17$\\
$T_d$		&$2315.8 \pm 1014.4$		&$1834.3 \pm 889.3$	&$3457.6 \pm 217.9$\\
$T_n$		&$1704.5 \pm 728.4$		&$1538.7 \pm 810.0$	&$1329.1 \pm 698.2$\\
$\sigma_K$	&$104.0 \pm 6.0$			&\nodata				&$105.0 \pm 6.0$\\
$\sigma_T$	&\nodata					&$107.0 \pm 7.0$		&$194.0 \pm 11.0$\\
$\log Z$		&$500.7 \pm 0.4$			&$1004.5 \pm 0.4$		&$950.6 \pm 0.2 $\\
$\log L_{\text{max}}$		&	$516.1$			&$1021.6$			&$973.3$		\\
				
\end{tabular}
\caption{Parameter estimates and log-evidences for the four models that were applied to the simulated \emph{Kepler} and TESS light curves.  For these simulations the photometric precision of the model-generated \emph{Kepler} and TESS data was assumed to be $60$ppm and $200$ppm, respectively.  }\label{tbl:results}
\end{table*}
Together, the forward model $F(t)$ can be written as the sum of all four components of the photometric flux
and the flux during transit and secondary eclipse, which is computed using the method of \citet{Mandel&Agol:2002}

\section{Results on Model-Generated Data}
\begin{figure}[h!]
\centering
\includegraphics[width = 8cm]{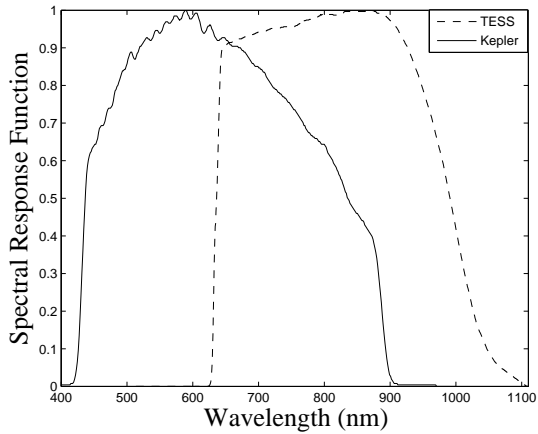}
\caption{The spectral response functions for both \emph{Kepler} (solid curve) and TESS (dashed curve).  Both are normalized so that their peaks occur at unity.  }\label{bandpasses}
\end{figure}

TESS will observe 26 sectors of the sky over a 2 year period and it will stare at each sector for approximately 27 days each.  It is expected that TESS will achieve a photometric precision of approximately 200 ppm \citep{Kraft+etal:2010}.  \emph{Kepler} on the other hand observed a patch of sky roughly $1/400^{th}$ the size that TESS will observe but achieved a photometric precision roughly an order of magnitude lower than TESS.  The observation sectors of TESS will overlap the \emph{Kepler} field and provide a unique opportunity to study a subset of the \emph{Kepler} planets in a slightly different bandpass.  
The \emph{Kepler} bandpass ranges from 420nm to 900nm whereas the TESS bandpass overlaps that of \emph{Kepler} ranging from 600nm to 1000nm.  The spectral response functions for both \emph{Kepler} and TESS are displayed in Figure \ref{bandpasses} with each function normalized to have a maximum of unity.  

Model-generated synthetic data was created for a hot-Jupiter in the \emph{Kepler} field observed by both \emph{Kepler} and TESS.  The synthetic \emph{Kepler} light curve spans a single quarter of observations ($\sim 90$ days), whereas the synthetic TESS light curve spans only 27 days corresponding to the duration that data will be taken from each observation sector.  The orbital parameters used to create the synthetic data for this planet were taken to be $e = 0$, $i = 84.3^o$ ($\cos i = 0.1$), $P = 1.7637$d, and $\omega = 0$, where $e,i,P$, and $\omega$ are the the orbital eccentricity, inclination, period, and argument of periastron, respectively.  Model parameters that describe the planetary characteristics were assumed to be $M_p = 7.5M_J$, $R_p = 1.4R_J$, $T_d = 3500$K, $T_n = 1500$K, and $A_g = 0.2$, where $M_p$ and $R_p$ are the planetary mass and radius, $T_d$ and $T_n$ are the day- and night-side temperatures of the planet, and $A_g$ is the geometric albedo of the planet.  Stellar parameters, which were held fixed, include the stellar mass $M_\star = 1.72M_\odot$, radius $R_\star = 1.71R_\odot$, and effective temperature $T_{\text{eff}} = 7650$K. 

\subsection{Model Selection}

A series of six simulations were conducted on the two data sets. The results of which are displayed in Table \ref{tbl:results}.  The goal was to determine whether or not one could disentangle thermal and reflected light by employing Bayesian model selection.  Using data from both \emph{Kepler} and TESS, the Bayesian log-evidence was computed for a model that included the reflection, Doppler beaming, and ellipsoidal variations effects but neglected thermal emission (labeled RBE in Table \ref{tbl:results}), and another model that included all four photometric effects (labeled RBET in Table \ref{tbl:results}).  For these simulations, the photometric precision of the simulated data from \emph{Kepler} and TESS were $60$ppm and $200$ppm, respectively. The two models were first applied to the simulated \emph{Kepler} and TESS data individually.  As shown in Table \ref{tbl:results}, the log-evidences for the RBE and RBET models were equal to within uncertainty. This indicates that with only a single dataset (TESS or \emph{Kepler} individually) one cannot disentangle reflected and thermally emitted light as expected.  Also note that the maximum log-likelihoods are equal when considering data from only a single instrument, meaning that both models yield the same overall best-fit. When considering data from both instruments simultaneously, the log-evidence corresponding to the RBE model was $\log Z = 949.0 \pm 0.2$ and for the RBET model it was determined to be $\log Z = 950.6 \pm 0.2$. This indicates that the model including thermal emission (RBET) was favored by a factor of approximately $\exp(1.6) \approx 5$ over the RBE model, which neglected thermal emission.  It should be noted that this planet was assumed to be in a circular orbit, which results in the thermal flux and reflected light variations to have approximately the same sinusoidal wave-form.  Observed through a single bandpass, this configuration is highly degenerate between the two effects.  This would result in the two equivalent log-evidences, or the RBE model being favored since the RBET model has a significantly higher penalty than the RBE model due to the addition of the day-and night-side temperature parameters.  For this simulation, the log-evidences were the same within uncertainty.  The maximum log-likelihood values also indicate that the RBET model is a better fit to the data than the RBE model. 

\begin{figure}[h!]
\centering
\includegraphics[scale = 0.56]{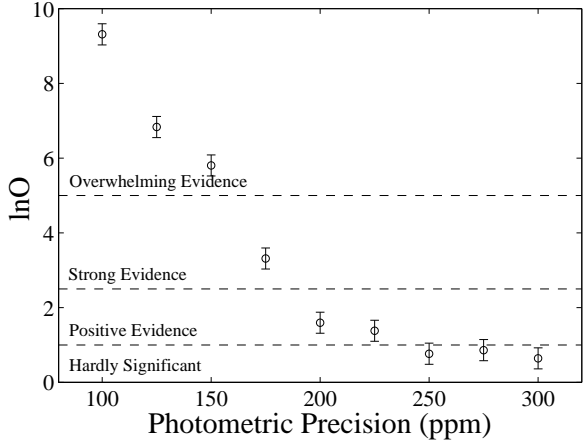}
\caption{Log-Bayes' factors for a series of nine simulations on model-generated \emph{Kepler} and TESS data.  While the photometric precision for the synthetic \emph{Kepler} data was held fixed at 100ppm, the photometric precision of the synthetic TESS data was increased from 100ppm to 300ppm in increments of 25ppm.  This plot shows that one should be able to disentangle thermal and reflected light if TESS can attain a photometric precision less than $\approx 175-225$ppm. }\label{log_Evidences}
\end{figure} 

Another experiment was performed to determine at what photometric precision TESS would need to achieve in order for the two signals to be distinguishable.  The RBE and RBET models were applied to nine pairs of simulated time series. The photometric precision of the model-generated \emph{Kepler} data was kept fixed at 60ppm, while the photometric precision for the simulated TESS observations were varied from 100ppm to 300ppm in steps of 25ppm.  Figure \ref{log_Evidences} displays the log-Bayes' factor for each pair of simulations.  Based on these results, thermal emission and reflected light should be definitively distinguishable for TESS observations that achieve a photometric precision of $< 200$ppm, while for noise levels greater than approximately 250ppm the two signals become indistinguishable once again.

\subsection{Parameter Estimation}
\begin{figure}[h!]
\centering
\includegraphics[scale = 0.28]{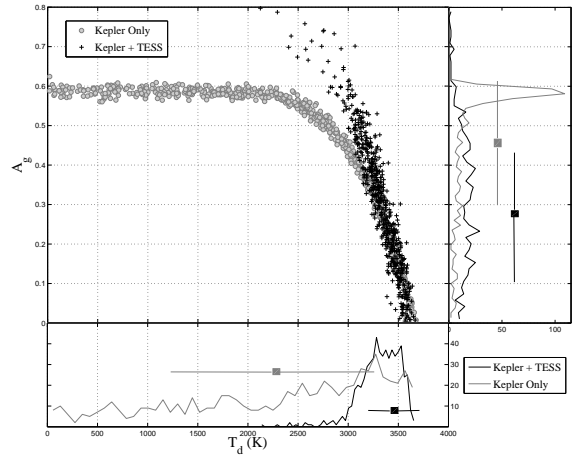}
\caption{The posterior samples for the RBET model applied to both the \emph{Kepler} and TESS data (black + signs), and the same model applied to only the \emph{Kepler} data (grey circles).  The one-dimensional histograms for the day-side temperature (B) and the geometric albedo (C) are also displayd.  Notice that the posterior for the dayside temperature, $T_d$, is much more peaked in the case of utilizing data from both bandpasses and that while the uncertainty in $A_g$ is hasn't improved by using \emph{Kepler} and TESS, the mean is much closer to the true value.  }\label{posteriors}
\end{figure}

The third aim of this study was to determine if one could better constrain either the day-side temperature or geometric albedo when using data from two separate bandpasses.  First, the RBET model was applied to the simulated \emph{Kepler} data alone to determine how well one could constrain these two parameters without TESS.  In this case, the day-side temperature and geometric albedo were found to be $T_d = 2315.8 \pm 1014.4$K and $A_g = 0.46 \pm 0.16$, respectively.  The uncertainty in the estimates of both of these parameters are large, especially in the case of the day-side temperature.  The reason for this can easily be seen in Figure \ref{posteriors} where the posterior samples for $T_d$ and $A_g$ are displayed as grey dots.  Note the signficant ridge-like degeneracy among these two parameters when the planet is observed through a single bandpass.
  This indicates that one could observe the same flux from a more reflective and cool planet, or a hot and dark planet.  Also displayed in Figure \ref{posteriors} are the posterior samples from the RBET model applied to both simulated \emph{Kepler} and TESS (200ppm) datasets (black $+$ signs).  In this case, the day-side temperature was much better constrained yielding  $T_d = 3457.6 \pm 217.9$K as indicated by the lack of posterior samples in the $0-2000$K range.  Even with two bandpasses it was not possible to precisely estimate the geometric albedo, which was estimated to be $A_g = 0.27 \pm 0.17$.  Although the uncertainty is still high, the mean is much closer to the true value of 0.2 since many of those samples corresponding to bright, cool objects could not explain the light curves.

\section{Doppler Beaming}
The amplitude of the observed beaming effect depends on the bandpass through which the planetary system is observed (equations \ref{beaming}-\ref{beaming_3}).  Therefore, it may be possible to further constrain the mass of exoplanets using more than one photometric channel.  Assuming a blackbody spectrum for the star, the photon-weighted bandpass-integrated beaming factor, $\alpha_b$, for the synthetic planet used above is  $\alpha_b = 3.27$ in the \emph{Kepler} bandpass.  Observed with TESS, the same factor is $\alpha_b = 2.64$ corresponding to an overall Doppler beaming amplitude change between bandpasses of $\approx 2$ ppm for a $7.5 M_{\text{jup}}$ planet.  

\begin{figure}[h!] \label{Beaming_Amplitudes}
\centering
\includegraphics[width = 8.1cm]{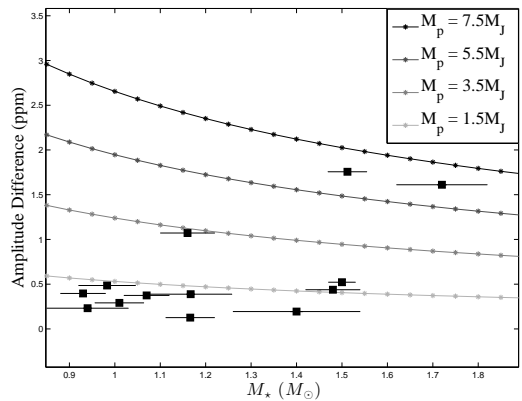}
\caption{Amplitude differences for the beaming effect observed with \emph{Kepler} and TESS. Each curve displays the $M_{\star}^{-2/3}$ dependence for Doppler beaming as shown in Eqn (\ref{semi-amplitude}). Also displayed (as squares) are the thirteen \emph{Kepler} hot-Jupiters likely to be detectable with TESS from Table \ref{Known Planets}.}
\end{figure}

With such a small difference even in the case of a massive planet, it is unlikely to significantly increase the accuracy or precision with which one can estimate the planetary mass. However, the amplitude difference also depends on the stellar mass as shown in Figure \ref{Beaming_Amplitudes}, which would make it possible to detect such a difference for large Hot-Jupiters around low-mass stars.  The difference in beaming amplitudes was calculated for each of the nine bright \emph{Kepler} hot-Jupiters for which there was a published planetary mass ($M_p \sin i$).  Figure \ref{Beaming_Amplitudes} displays the amplitude difference between \emph{Kepler} and TESS for each planetary system (shown as black squares) along with four synthetic planets around stars with masses varying from $0.1M_\odot$ to $3.0 M_\odot$.  Each of the systems yields amplitude differences less than $2$ppm, which is unlikely to significantly increase the level of exoplanet characterization given the expected photometric precision of TESS.

\section{Discussion}
Using both Bayesian model selection and parameter estimation we have shown that TESS should be able to further characterize a subset of the \emph{Kepler} planets when it's observation sector overlaps the \emph{Kepler} field.  By employing two sets of simulated observations that were taken through different bandpasses, one can disentangle thermal emission and reflected light originating from the atmosphere of the planet.  This can be done even in the case of circular orbits, which yield the same flux variations for the two effects.  Tests on simulated data indicate that if TESS can achieve a photometric precision less than approximately 250ppm for planets in the \emph{Kepler} field, then the thermal and reflected flux can be disentangled.  

It was also shown that by incorporating more than just a single bandpass, the degeneracy among the day-side temperature and the geometric albedo of the planets decrease allowing for more accurate characterization of such planets.  In the case of the \emph{Kepler} bandpass alone, the estimated day-side temperature of a simulated short-period hot Jupiter was found to be $T_d = 2315.8 \pm 1014.4$K with a geometric albedo of $A_g = 0.46 \pm 0.16$.  Using both bandpasses yielded a day-side temperature of $T_d = 3457.6 \pm 217.9$K and geometric albedo of $A_g = 0.27 \pm 0.17$, which are much closer to the true values of $T_d = 3500$K and $A_g = 0.2$. 

Time series obtained from the TESS mission that overlaps with the \emph{Kepler} field will allow for more precise characterization of the brightest \emph{Kepler} systems, and may even shed light on current planetary candidates.  Although this study was specific to \emph{Kepler} and TESS, the methodology is applicable to any multi-channel  (HST, JWST, etc.) exoplanet observations and may be utilized to plan future missions and mission target lists with the goal of exoplanet characterization.

\section{Acknowledgements}
This research has made use of the Exoplanet Orbit Database and the Exoplanet Data Explorer at exoplanets.org.

%
\bibliographystyle{plainnat}
\bibliography{bibliography}

\end{document}